\documentclass[superscriptaddress, reprint, amsmath, amssymb, aps, pra, floatfix]{revtex4-2}

\usepackage[utf8]{inputenc}
\usepackage{graphicx}% Include figure files
\usepackage{dcolumn}% Align table columns on decimal point
\usepackage{bm}% bold math
\usepackage[normalem]{ulem} % For strikethrough
\usepackage{mathrsfs}
\usepackage{physics}
\usepackage{xcolor}
\usepackage{amsmath, amssymb, mathtools, amsthm, siunitx}
\usepackage[colorlinks=true]{hyperref}
\usepackage{comment} %create block comment
\usepackage{orcidlink}

\usepackage{xr}
\definecolor{ve}{RGB}{0,161,22}

\begin{abstract}
Quantum tracking control encodes the desired dynamics into a tailored driving field; here, we let the system find its own way there. We propose a real-time feedback control framework in which a proportional controller continuously corrects a simple transform-limited field based on the instantaneous mismatch between two systems' responses---producing the required control on the fly, without prior waveform design. The framework is demonstrated on two distinct examples: a single-active-electron atom, where hydrogen is driven to mimic argon's strong-field optical emission, and a Fermi–Hubbard chain, where a weakly interacting lattice reproduces the transport dynamics of a Mott-insulating reference. By shifting the control paradigm from predesigned inputs to adaptive response tracking, this approach establishes closed-loop feedback as a broadly applicable route to programmable quantum dynamics.
\end{abstract}

\begin{document}
\author{Valeriia Bilokon\orcidlink{0009-0001-1891-0171}}
\email{vbilokon@tulane.edu}
\affiliation{Department of Physics and Engineering Physics, Tulane University, New Orleans, Louisiana 70118, United States}
\affiliation{Akhiezer Institute for Theoretical Physics, NSC KIPT, Akademichna 1, 61108 Kharkiv, Ukraine}

\author{Elvira Bilokon\orcidlink{0009-0007-8296-2906}}
\email{ebilokon@tulane.edu}
\affiliation{Department of Physics and Engineering Physics, Tulane University, New Orleans, Louisiana 70118, United States}
\affiliation{Akhiezer Institute for Theoretical Physics, NSC KIPT, Akademichna 1, 61108 Kharkiv, Ukraine}

\author{Denys I. Bondar \orcidlink{0000-0002-3626-4804}}
\email{dbondar@tulane.edu}
\affiliation{Department of Physics and Engineering Physics, Tulane University, New Orleans, Louisiana 70118, United States}

\title{All You Need is Amplifier: Spectral Imposters Without Pulse Shaping}

\maketitle

%% ========================================================================
%  Introduction
%% ========================================================================
\textit{Introduction---}Quantum materials host a wealth of emergent phenomena---strong electronic correlations~\cite{Imada1998, Lee2006}, topology~\cite{Hasan2010, Qi2011, Keimer2017}, unconventional magnetism~\cite{Balents2010, Savary2017}, and extreme nonlinear optical responses~\cite{lewenstein_theory_1994, Krausz2009, Ghimire2011, Uchida2022, Murakami2022}---that remain among the most compelling targets of modern physics, with far-reaching implications for quantum technology and materials design~\cite{Basov2017, Torre2021}. Yet the most exotic behaviors often arise in systems that are difficult to access or manipulate: they may require extreme conditions~\cite{Fausti2011, Drozdov2015}, exhibit limited tunability~\cite{Oka2019, Abanin2019, Kobayashi2023}, or have parameters fixed by chemistry and structure~\cite{Imada1998, Kotliar2004, Keimer2015}. This  motivates a broader aspiration: rather than synthesizing a new material for every desired functionality, can we make an accessible and controllable system reproduce the dynamical response of a less accessible one~\cite{Bloch2008, Campos2017, McCaul2020, Maskara2025, Lakes2025}? In this view, the central goal shifts from discovering phases to programming responses~\cite{Judson1992, Rabitz2000}.

A common route toward such response engineering relies on open-loop optimal control and pulse shaping~\cite{Peirce1988, Werschnik2007, Brif2010, Mitra2024}. In that paradigm, one computes a tailored driving field designed to coerce the system into matching a predefined target signal~\cite{Zhu1998, Koch2022}. Closed-loop learning control~\cite{Judson1992, Yelin1997, Bardeen1997, Assion1998, Levis2001} in ultrafast optics demonstrated that complex quantum dynamics can indeed be steered through carefully optimized waveforms. More recent advances extend these ideas to condensed-matter settings, where shaped pulses are used to sculpt nonlinear optical and transport responses~\cite{Kunde2000, Campos2017, Magann2018, Schaefer2020, McCaul2020, Mccaul2021, Shan2021, Lange2024, Richter2024}. While powerful, this strategy places the burden on the input design: it demands detailed system knowledge or iterative optimization and often results in intricate, system-specific fields.

In this Letter, we propose a radically simpler paradigm: all you need is amplifier. Rather than encoding target dynamics into complex-shaped pulses, we drive the system with a simple transform-limited field and generate the necessary control dynamically through optical feedback. The key hardware requirement is not a pulse shaper, but an amplifier---a standard workhorse of optics. We demonstrate the generality of the framework on two physically distinct platforms: a single atom in the strong-field and an interacting many-body lattice described by the Fermi–Hubbard model. In both cases, control is realized not by sculpting the input waveform, but by embedding the system within an active amplification loop that dynamically reshapes the effective driving through feedback.

Coupling a quantum system to such an amplification channel has recently emerged as a powerful route to engineer effective Hamiltonians, stabilize otherwise inaccessible dynamical regimes, and induce novel nonequilibrium phases across condensed-matter and photonic platforms~\cite{Yao2023, Rasputnyi2024, Wang2025}. In particular, feedback-mediated amplification has enabled phenomena such as epsilon-near-zero (ENZ) response generation~\cite{Masur2023} that inspired a recent experimental demonstration of light-induced ENZ behavior~\cite{Runge2025}.
Viewed in this broader context, our results establish amplification as the essential hardware resource for enforcing target quantum responses. By replacing waveform design with real-time gain, the framework offers a broadly applicable route toward programmable quantum dynamics using only the simplest optical inputs.

%% ========================================================================
%  Why Intensity?
%% ========================================================================
\textit{Why intensity?---}Since an optical amplifier increases intensity, we begin by outlining the physical intuition for why intensity modulation is sufficient to realize driven imposters, i.e., to match the optical responses of two distinct systems.
When an atom is irradiated by a strong near-infrared laser field, it emits photons at frequencies that are odd multiples of the frequency of the incident light $\omega$. This phenomenon is known as high-harmonic generation (HHG) and underlies attosecond science and technology~\cite{Krausz2009}. The highest frequency emitted, i.e., the HHG cutoff, is given by
$
    \Omega_{\max} = 3.17\, U_p + I_p
$~\cite{Kulander1992, Corkum1994, lewenstein_theory_1994} (atomic units are used throughout unless stated otherwise), where $U_p = (F / (2\omega))^2$ is the ponderomotive potential, $F$ is the strength of the laser field, and $I_p$ is the atom's ionization potential. We take this HHG spectrum as a reference benchmark. If we want to find a laser pulse that induces the same reference HHG spectrum but from a different atom with ionization potential $I_p'$, the simplest approach is to use the same laser frequency but adjust the laser field strength to
\begin{align}
    F' = 1.12\, \omega \sqrt{\Omega_{\max} - I_p'}.
\end{align}
This choice guarantees that the HHG peaks occur at the same odd-multiple frequencies of $\omega$ and have the identical cutoff frequency $\Omega_{\max}$ as the reference spectrum. Hence, we can achieve similar nonlinear optical responses by using a simple monochromatic laser field of appropriately chosen intensity, without the need to resort to pulse shaping~\cite{Campos2017}.

This mimicry is not restricted to HHG; it also applies to the photoelectron spectrum. In above-threshold ionization (ATI), an electron is ejected from the atom with kinetic energy $p^2/2 = n \omega - U_p - I_p$~\cite{faisal1973multiple, Reiss_1980} by absorbing $n$ photons. The peaks of ATI photoelectron spectra correspond to the number of absorbed photons. To induce a similar photoelectron spectrum from an atom with a different ionization potential $I_p'$, we should keep the frequency unchanged and choose the new field strength 
\begin{align}
    F' = 2\omega \sqrt{U_p + I_p - I_p'},    
\end{align}
such that $U_p + I_p = U_p' + I_p'$, where $U_p' = (F'/(2\omega))^2$. Thus, two atoms achieve identical photoelectron emission.

%% ========================================================================
%  Feedback Control General Part
%% ========================================================================
\textit{Feedback Control Framework---}We consider two quantum systems: a reference system described by a Hamiltonian $\mathcal{\hat{H}}_{\rm ref}(t)$ driven by a simple transform-limited field $E_{\rm tl}(t)$, and a driven system governed by a Hamiltonian $\mathcal{\hat{H}}_{\rm dr}(t)$. Specifically, the  field $E_{\rm tl}(t)$ is a monochromatic pulse with a $\sin^2$ envelope,
\begin{equation}
    E_{\rm tl}(t) = E_0 \cos(\omega_0 t)\sin^2\!\left(\frac{\pi t}{T}\right), \quad 0 \le t \le T,
\end{equation}
where $E_0$ is an amplitude of the laser field strength, $\omega_0$ is a laser field frequency, and $T$ is the total pulse duration ($T=2\pi N/\omega_0$, where $N$ is the number of periods). The reference system produces a response $Y(t) = \frac{d}{dt}\langle \hat{O} \rangle_{\rm ref}$ when irradiated by $E_{\rm tl}(t)$, which serves as the target signal. Our objective is to design a control protocol that forces the driven system to reproduce this response, i.e., $\frac{d}{dt}\langle \hat{O} \rangle_{\rm dr} \to Y(t)$, despite differences in the Hamiltonians $\mathcal{\hat{H}}_{\rm ref}$ and $\mathcal{\hat{H}}_{\rm dr}$.

\begin{figure}
    \centering
    \includegraphics[width=0.9\linewidth]{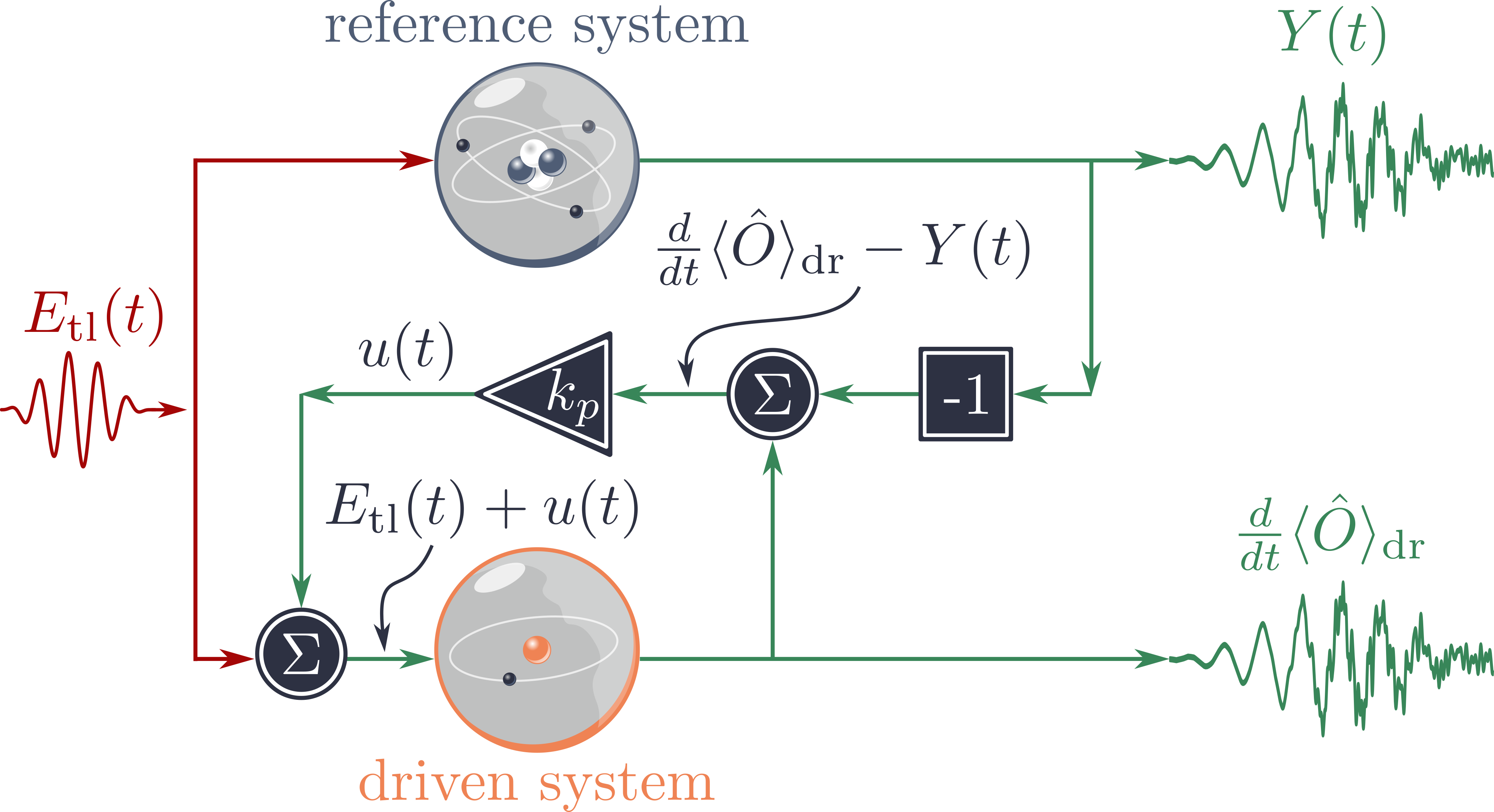}
    \caption{Driven imposters via optical feedback. A? proportional controller feeds back the optical response of the driven system until the generated field reproduces the response of the reference system.}
    \label{fig:schematic_setup}
\end{figure}

The feedback architecture, shown schematically in Fig.~\ref{fig:schematic_setup}, implements a proportional control loop. Both systems are driven by the same transform-limited field $E_{\rm tl}(t)$, while the driven system experiences an additional feedback correction $u(t)$, so that its total input field is $E_{\rm dr}(t)=E_{\rm tl}(t)+u(t)$. The feedback signal is generated from the instantaneous mismatch between the system responses according to
\begin{equation}
u(t) = k_p \left[ \frac{d}{dt}\langle \hat{O} \rangle_{\rm dr} - Y(t) \right],
\label{eq:proportional_control}
\end{equation}
where $k_p$ is the proportional gain achieved by an amplifier.

The dynamics of the observable are governed by the Ehrenfest theorem
\begin{equation}
    \frac{d}{dt}\langle \hat{O} \rangle = \left\langle \frac{\partial \hat{O}}{\partial t} \right\rangle + i\langle [\mathcal{\hat{H}}, \hat{O}] \rangle.
\label{eq:ehrenfest}
\end{equation}
Substituting Eq.~\eqref{eq:ehrenfest} into the control law~\eqref{eq:proportional_control} yields a self-consistent equation for the feedback signal,
\begin{equation}
    u = k_p \left[\left\langle \frac{\partial \hat{O}}{\partial t} \right\rangle_{\rm dr}
    + i\langle [\mathcal{\hat{H}}_{\rm dr},\hat{O}] \rangle_{\rm dr} - Y \right].
\label{eq:control_field}
\end{equation}
Note that $u$ appears implicitly on the right-hand side of Eq.~\eqref{eq:control_field}.

Until now, the discussion has been general. The remaining ingredient is to specify $\mathcal{\hat{H}}_{\rm dr}$ and $\hat{O}$. In both cases studied below, in the high gain limit $k_p\to\infty$, the feedback loop enforces the tracking condition $\frac{d}{dt}\langle \hat{O} \rangle_{\rm dr} \to Y(t)$. In practice, finite values of $k_p$ are sufficient to achieve stable tracking, as demonstrated in the following examples.

%% ========================================================================
%  Single-particle Control
%% ========================================================================
\textit{Teaching Hydrogen to Mimic Argon---}Consider a single-active-electron model~\cite{Bauer1997} of an atom driven by a linearly polarized laser pulse. The field-free Hamiltonian is
\begin{equation}
\mathcal{\hat{H}}_{\rm at}^{(0)}=\hat{p}^2 / 2+V_{\rm at}(\hat{x}),
\end{equation}
where $\hat{p}$ denotes a momentum operator for the electron and $V_{\rm at}(\hat{x})$ is a soft-Coulomb effective potential modeling the electron--core interaction. For simplicity, we restrict ourselves to the 1D case, although the statements below remain valid in 3D and for an arbitrary number of electrons. 

The reference dynamics is generated by irradiating the atom with a simple transform-limited pulse $E_{\rm tl}(t)$ in the length gauge,
$
\mathcal{\hat{H}}_{\rm ref}(t)=\mathcal{\hat{H}}_{\rm at}^{(0)}+\hat{x}\,E_{\rm tl}(t).
$
According to the Larmor formula, the electromagnetic field emitted by an active electron is proportional to its acceleration, $d\langle \hat{p}\rangle/dt$~\cite{Brabec2000}. We therefore choose the electronic momentum as the tracking observable, $\hat{O}\equiv\hat{p}$, and define the reference signal as $Y(t)\equiv \frac{d}{dt}\langle \hat{p}\rangle_{\rm ref}$.

We now introduce the control channel by allowing the driven system to experience, in addition to the transform-limited input, a feedback-generated correction field $u(t)$, 
$
\mathcal{\hat{H}}_{\rm dr}(t)=\mathcal{\hat{H}}_{\rm at}^{(0)}+\hat{x}[E_{\rm tl}(t) + u(t)]=\mathcal{\hat{H}}_{\rm at}^{(0)}+\hat{x}E_{\rm dr}(t).
$
The proportional controller first subtracts the two optical responses (this can be done, e.g., via a 50/50 beam splitter) and then amplifies the resulting signal; hence, the output from the optical amplifier is 
$
u(t)=k_p\!\left[\frac{d}{dt}\langle \hat{p}\rangle_{\rm dr}-Y(t)\right].
$
Using Ehrenfest’s theorem~\eqref{eq:ehrenfest}, the feedback signal can be explicitly written in closed form. For the atomic Hamiltonian in the length gauge, the driven response obeys $\frac{d}{dt}\langle \hat{p}\rangle_{\rm dr}
= \langle \hat{F} \rangle_{\rm dr} - E_{\rm dr}(t)$, where $\hat{F} = -V_{\rm at}'(\hat{x})$ models the interaction of the electron with a core~\cite{Campos2017}. As a result, the control field yields
\begin{equation}
u(t)=\frac{k_p}{1+k_p}\left[\langle \hat{F}\rangle_{\rm dr}-E_{\rm tl}(t)-Y(t)\right],
\label{eq:control_field_atom}
\end{equation}
and in the large-gain regime, the driven response is forced to follow the reference, $\frac{d}{dt}\langle \hat{p}\rangle_{\rm dr}\to Y(t)$.

\begin{figure}
        \centering
        \includegraphics[width=1\linewidth]{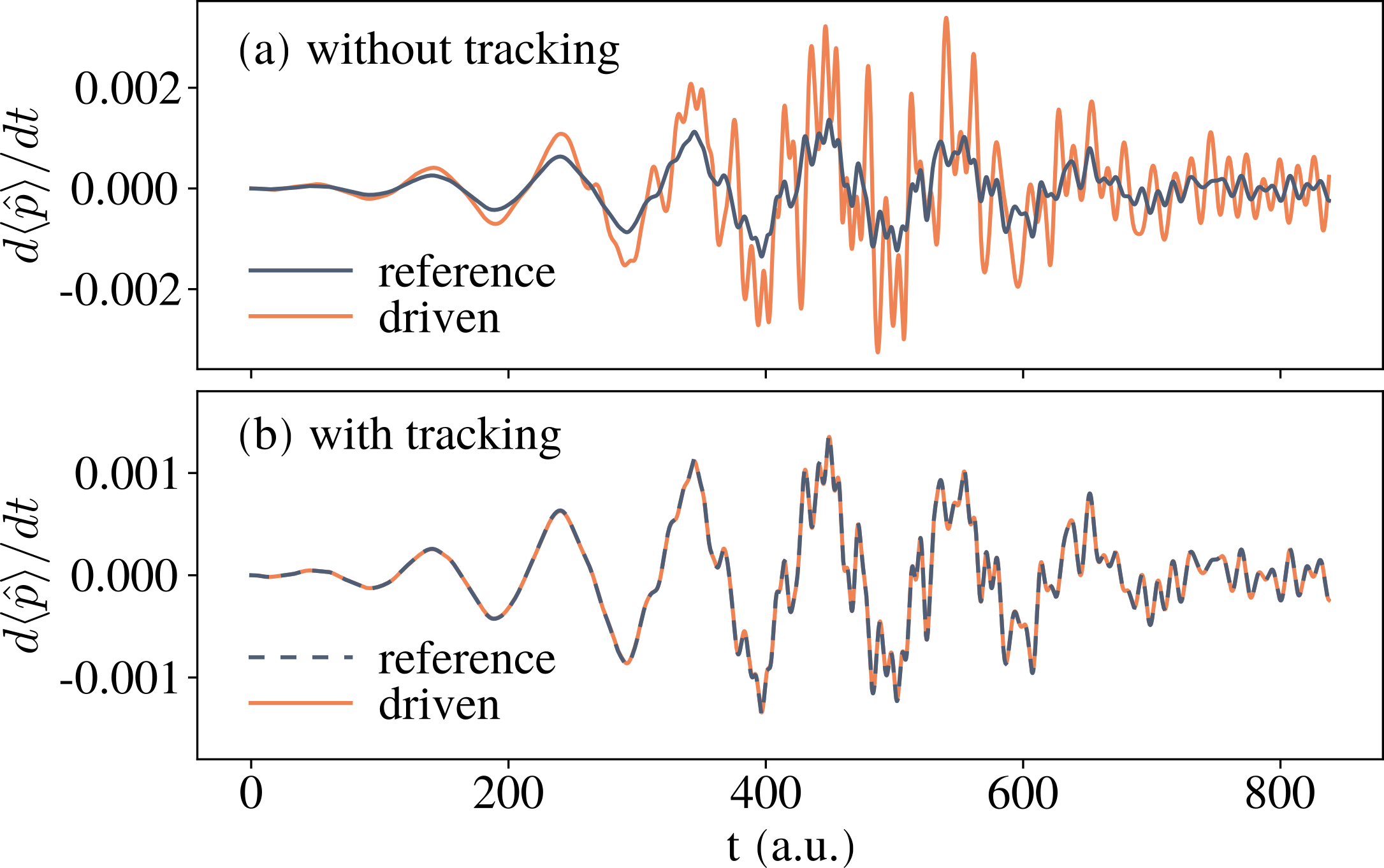}
        \caption{Time-domain high-harmonic response $d\langle \hat{p}\rangle/dt$ for argon (reference, blue) and hydrogen (driven, orange) within the single-active-electron model. (a) Without feedback: the two atoms exhibit distinct nonlinear responses under identical transform-limited driving $E_{\rm tl}(t)$. (b) With amplification gain $k_p=1000$: the hydrogen response converges to the argon reference signal, demonstrating dynamical tracking and optical mimicry.}
        \label{fig:AtomFeedback}
\end{figure}

To demonstrate the tracking principle in a concrete setting, we consider two distinct atomic systems within the single-active-electron approximation: argon as the reference and hydrogen as the driven system. Figure~\ref{fig:AtomFeedback}(a) shows the $d\langle \hat{p}\rangle/dt$ response for both atoms in the absence of feedback. The two signals clearly exhibit distinct amplitudes and temporal structures, reflecting the different underlying electronic dynamics.

We now activate the proportional feedback loop introduced above, with amplification $k_p=1000$, allowing the control channel to dynamically modify the driving field based on the instantaneous deviation from the argon response. Figure~\ref{fig:AtomFeedback}(b) demonstrates that under feedback control, the hydrogen response accurately tracks the argon reference signal. The feedback-generated field $u(t)$ defined in Eq.~\eqref{eq:control_field_atom} effectively compensates for the difference in atomic potentials, enabling hydrogen to mimic argon's nonlinear optical behavior.

This result illustrates that feedback control can dynamically compensate intrinsic atomic differences and enforce dynamical equivalence between otherwise distinct quantum systems even in an extremely challenging case of HHG. In the high-gain limit, the driven system becomes dynamically indistinguishable from the reference at the level of its optical response.

%% ========================================================================
%  Many-body Control
%% ========================================================================
\textit{Engineering Many-Body Interactions on Demand---}
We now extend the tracking-control framework to an interacting quantum many-body system. As an example, we consider a one-dimensional Fermi--Hubbard model, described by the following Hamiltonian: 
\begin{align}
\mathcal{\hat{H}}(t) = &- t_0 \sum_{j,\sigma}
\left( e^{-i\Phi(t)} \hat{c}^\dagger_{j,\sigma} \hat{c}_{j+1,\sigma}
+ e^{i\Phi(t)} \hat{c}^\dagger_{j+1,\sigma} \hat{c}_{j,\sigma} \right) \nonumber \\
&+ U \sum_j \hat{n}_{j\uparrow}\hat{n}_{j\downarrow},
\label{eq:fhm_peierls}
\end{align}
where $\hat{c}^\dagger_{j,\sigma}$ ($\hat{c}_{j,\sigma}$) creates (annihilates) a fermion with spin $\sigma\in\{\uparrow,\downarrow\}$ at lattice site $j$, and $\hat{n}_{j\sigma}=\hat{c}^\dagger_{j,\sigma}\hat{c}_{j,\sigma}$ is the corresponding number operator. The hopping amplitude is denoted by $t_0$, $U$ is the on-site interaction strength, and $\Phi(t)$ is a time-dependent Peierls phase~\cite{Peierls1933, Nocera2017} encoding the coupling to an external electric field via
$
E(t) = -\frac{1}{a}\frac{d\Phi(t)}{dt},
$
with the lattice spacing $a$.

The observable of interest is the many-body current operator,
\begin{equation}
\hat{J}(t) = - i a t_0 \sum_{j,\sigma}
\left(e^{-i\Phi(t)} \hat{c}^\dagger_{j,\sigma} \hat{c}_{j+1,\sigma}
-e^{i\Phi(t)} \hat{c}^\dagger_{j+1,\sigma} \hat{c}_{j,\sigma}\right),
\end{equation}
which plays a role analogous to the electronic momentum in the single-particle setting. Here and below we use units $\hbar=1$. Accordingly, we define the target signal as the time derivative of its expectation value, 
$
Y(t)\equiv \frac{d}{dt}\langle \hat{J} \rangle_{\rm ref},
$
generated by a reference system driven solely by the transform-limited field $E_{\rm tl}(t)$. 

The driven system is irradiated by the combined field $E_{\rm tl}(t)+u(t)$ and placed into a proportional feedback loop that compares its response to the target signal, yielding
$
u(t)
=
k_p
\left[
\frac{d}{dt}\langle \hat{J} \rangle_{\rm dr}
-
Y(t)
\right].
$
Using Ehrenfest’s theorem~\eqref{eq:ehrenfest}, the feedback signal can be written explicitly in closed form. The resulting self-consistent control law reads
\begin{align}
u(t)
=
\frac{k_p}{1 + a k_p \langle \hat{\mathcal{H}}_{\rm kin} \rangle_{\rm dr}}
\Big[
&- a E_{\rm tl}(t)\langle \hat{\mathcal{H}}_{\rm kin} \rangle_{\rm dr}
\notag\\
&+\, i\langle [\hat{\mathcal{H}},\hat{J}] \rangle_{\rm dr}
- Y(t)
\Big].
\label{eq:control_field_fhm}
\end{align}
and in the large-gain limit
$
\frac{d}{dt}\langle \hat{J} \rangle_{\rm dr}\to Y(t).
$ 
Note that the feedback channel becomes singular as $\langle \hat{\mathcal{H}}_{\rm kin} \rangle_{\rm dr}\to0$, signaling a loss of controllability when the current no longer responds to Peierls-phase variations.

\begin{figure}[h!]
        \centering
        \includegraphics[width=1\linewidth]{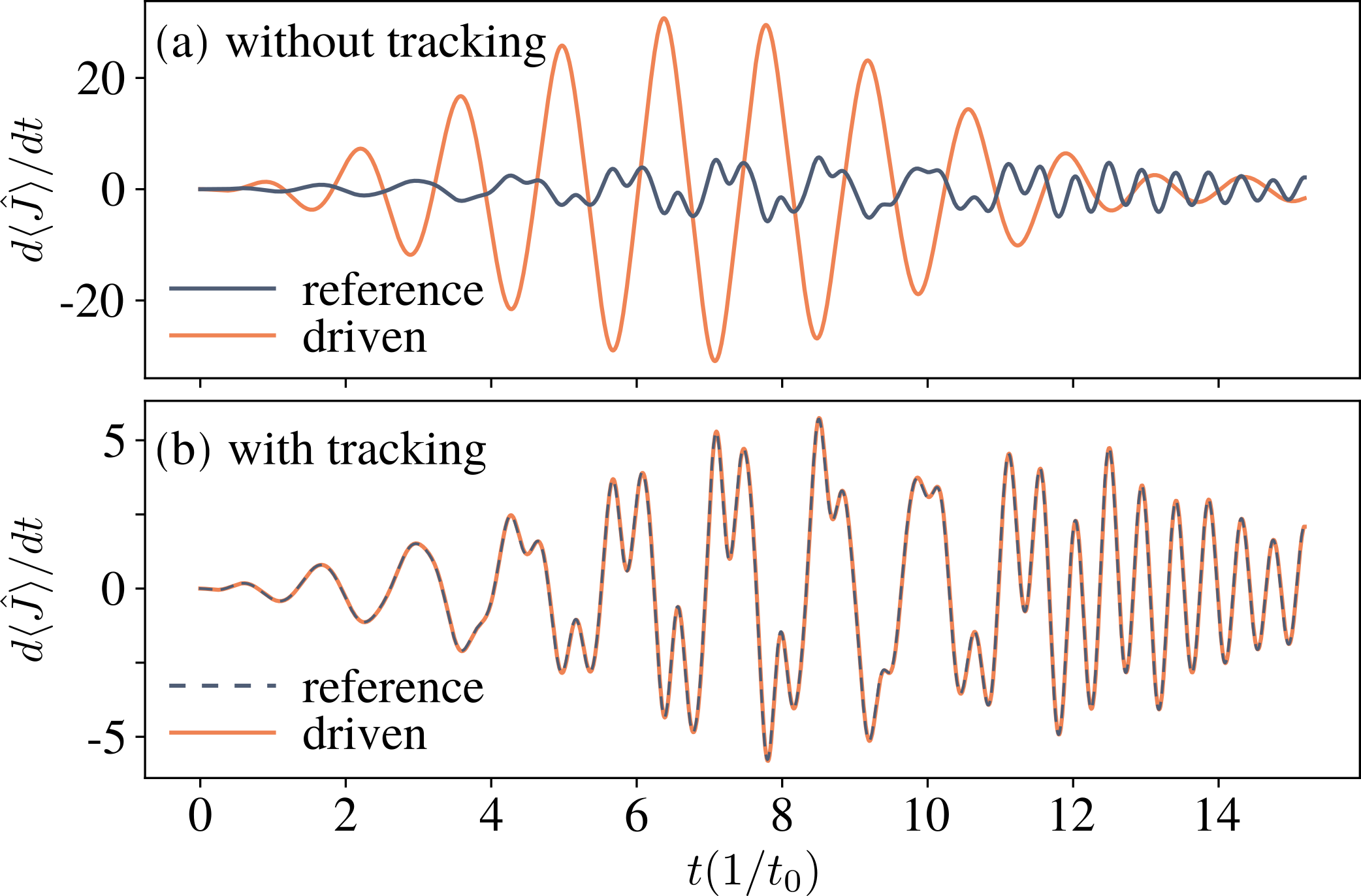}
        \caption{Many-body current tracking in the one-dimensional Fermi--Hubbard model ($L=10$, half filling), $U_{\rm ref}/t_0=10$ and $U_{\rm dr}/t_0=1$. (a) Without feedback, the time derivative of the current expectation value differs significantly in amplitude and phase. (b) With amplifier gain $k_p=1000$, the driven system reproduces the reference response, $\frac{d}{dt}\langle \hat{J} \rangle_{\rm dr} \to \frac{d}{dt}\langle \hat{J} \rangle_{\rm ref}$.}
        \label{fig:HubbardFeedback}
\end{figure}

To demonstrate many-body tracking, we consider a one-dimensional Fermi–Hubbard chain of $L=10$ sites at half filling with periodic boundary conditions. The reference system is chosen to be strongly correlated, with interaction strength $U_{\rm ref}/t_0=10$, corresponding to a Mott-insulating regime, while the driven system is taken to be weakly interacting, $U_{\rm dr}/t_0=1$. We set the hopping amplitude to $t_0=0.35$~eV and the lattice spacing to $a=\SI{3.8}{\angstrom}$~\cite{Scalapino2012, Andersen1995}. Both systems are irradiated by a transform-limited pulse of duration $N=10$ optical cycles with frequency $\omega_0=375$~THz and field amplitude $E_0=24$~MV/cm, consistent with experimentally demonstrated strong-field conditions in solids~\cite{Kim2017}.

Without feedback, the current dynamics differ substantially due to the distinct interaction strengths. These differences are clearly visible in Fig.~\ref{fig:HubbardFeedback}(a), where $\frac{d}{dt}\langle \hat{J} \rangle$ differs both in amplitude and phase. When the proportional feedback loop is activated, the control field $u(t)$ modifies the Peierls phase according to Eq.~\eqref{eq:control_field_fhm}. As shown in Fig.~\ref{fig:HubbardFeedback}(b), for finite gain $k_p=1000$, the driven system converges to the reference one, i.e., $\frac{d}{dt}\langle \hat{J} \rangle_{\rm dr} \to \frac{d}{dt}\langle \hat{J} \rangle_{\rm ref}$. 

Thus, a weakly interacting system can be driven to reproduce the transport response of a strongly correlated Mott regime. In this sense, the feedback field dynamically compensates for the missing correlations in the weakly interacting system.

%% ========================================================================
%  Conclusions
%% ========================================================================
\textit{Conclusions---}We have demonstrated that feedback-driven amplification can be used to enforce target quantum responses across physically distinct systems even for an extreme optical nonlinearities such as HHG. The framework is validated on two complementary platforms: hydrogen reproducing argon’s strong-field optical emission, and a weakly interacting Fermi--Hubbard chain replicating the transport dynamics of a Mott-insulating reference. Together, these examples establish the generality of response-level control from single-particle to correlated many-body regimes.

Apart from the proven capability of generating ENZ optical behavior via feedback-mediated amplification~\cite{Masur2023}, the present framework suggests a direct route toward controlling collective quantum phases. In particular, feedback-based gain control could be applied to enhance or suppress superconducting correlations, which have previously been manipulated using carefully tailored pulse sequences~\cite{Povitchan2025}. By replacing intricate pulse shaping with real-time adaptive correction, the approach provides a broadly applicable route to shaping quantum dynamics using minimal input structure.

\textit{Acknowledgments.} 
This work was supported by Army Research Office (ARO) (grant W911NF-23-1-0288; program manager Dr.~James Joseph).
E.B. was supported by the National Science Foundation (NSF) IMPRESS-U Grant No.~2403609. The views and conclusions contained in this document are those of the authors and should not be interpreted as representing the official policies, either expressed or implied, of ARO, NSF, or the U.S. Government. The U.S. Government is authorized to reproduce and distribute reprints for Government purposes notwithstanding any copyright notation herein.

\bibliography{main}

\end{document}